\documentclass[aps,twocolumn,showpacs,groupedaddress]{revtex4}  
\usepackage{graphicx}  
\usepackage{dcolumn}   
\usepackage{bm}        
\usepackage{amssymb}   

\begin{document}


\title{Heisenberg limited metrology using Quantum Error-Correction Codes}
\author{Roee~Ozeri}
\affiliation{Department of Physics of Complex Systems, Weizmann Institute of
Science, Rehovot 76100, Israel}

\begin{abstract}
Methods borrowed from the world of quantum information processing have lately been used to enhance the signal-to-noise ratio of quantum detectors. Here we analyze the use of stabilizer quantum error-correction codes for the purpose of signal detection. We show that using quantum error-correction codes a small signal can be measured with Heisenberg limited uncertainty even in the presence of noise. We analyze the limitations to the measurement of signals of interest and discuss two simple examples. The possibility of long coherence times, combined with their Heisenberg limited sensitivity to certain signals, pose quantum error-correction codes as a promising detection scheme.
\end{abstract}

\maketitle

The evolution of quantum superpositions under the influence of external fields can be used for the measurement of the latter. In particular, two-level quantum systems (pseudo-spins or qubits) have been used for the measurement of electric \cite{Osterwalder1999,Dolde2011}, magnetic \cite{BudkerRomalis2007, Maze2008} and gravitational \cite{KasevichChu1991, Chou2010} fields among other examples. The precision with which such measurements are preformed is ultimately limited by noise. In the absence of external noise the uncertainty with which the rotation frequency of a spin-superposition can be estimated is limited to $\Delta\xi \geq \frac{1}{2\tau\sqrt{N}}$ where $\tau$ is the rotation time and $N$ is the number of times the experiment was independently repeated (either by sequential repetitions using the same probe or by parallel measurements on many uncorrelated probes). This limitation to precision is known as the Standard Quantum Limit (SQL) and is the result of quantum projection noise. Given a fixed total time of data acquisition, it pays-off to perform longer experiments with fewer repetitions rather than repeating shorter experiments a for larger number of times. This situation changes in the presence of external noise leading to probe decoherence. Here, whenever an experiment is longer than the typical decoherence time, the information on the superposition rotation angle is lost. The optimal experiment duration will be roughly the decoherence time \cite{Huelga1997}. To be able to measure a small signal it is desirable to extend the time during which the probe superposition coherently evolves.


The failure of maintaining superpositions coherent is an obstacle for the realization of many quantum technologies. In particular in the field of quantum information processing, decoherence of quantum superpositions renders quantum computers classical. Active decoherence prevention methods were therefore developed in this context. Examples include, dynamic decoupling protocols, which were originally used in nuclear magnetic resonance experiments \cite{Hahn1950, CarrParcell1954, MeiboomGill1958, Gordon2007}, the encoding of information in decoherence-free subspaces, which are spanned by the common degenerate eigenstates of the noise operators \cite{Lidar1998}, and quantum error-correction codes \cite{Shor1995, Steane1996}. In the latter, the operation of noise operators can rotate states to outside, but not in, the code subspace. Repetitive measurements of the subspace in which the superposition resides followed by re-encoding in the code subspace protects against noise. It was shown that by concatenating layers of some quantum error-correction codes, a quantum register of qubits can be made fault-tolerant and maintain encoded superpositions coherent indefinitely \cite{Aharonov1998}.

The long coherence times offered by some of the methods above have been used for the purpose of reducing the uncertainty of measurements. The signals that can be measured have to act as gates on the protected probe, i.e. rotate the protected probe superposition. Correspondingly, dynamic decoupling methods have been used to measure alternating magnetic fields and light shifts with pristine sensitivity \cite{Maze2008, Kotler2011}. Decoherence-free subspaces, that are resilient to the effect of collective magnetic field noise, were used for the measurement of magnetic field gradients and magnetic dipolar coupling between electronic spins in trapped atomic ions \cite{Haeffner2005, Kotler2013}.

Here we propose the use of Quantum Error-Correction Codes (QECC's) to reduce the uncertainty of measurements in the presence of noise. Similarly to the two previous examples of using dynamic decoupling and decoherence-free subspaces to this end, QECC's efficiently protect against certain noise operators, whereas other (sometimes highly correlated) operators act as single qubit rotations in the QECC. In the context of quantum computing, these operators are used in order to perform logical gates in the code. In the context of quantum metrology, the strength with which these operators couple to the encoded qubit can be measured by estimating its rotation angle. Since the dynamics of a qubit encoded in the QECC is highly protected against noise, its coherent evolution can be, in principle, unperturbed for very long times, allowing for the measurement of very weak signals. We analyze the use of stabilizer QECC for metrology and show that, not only is the measurement of weak signals possible in the presence of noise, but the measurement uncertainty can be Heisenberg limited in the number of qubits participating in the code.

As an example, consider the three-qubit superpositions which are the basic building block in Shor's nine-qubit repetition code. Here, a single logical qubit is encoded in the subspace spanned by $|\uparrow\uparrow\uparrow\rangle$ and $|\downarrow\downarrow\downarrow\rangle$. This code corrects against erroneous single qubit flips, $XII$, $IXI$ and $IIX$. Here $X$, $Y$ and $Z$ are the corresponding spin $1/2$ Pauli operators, and the order by which they appear indicates the qubit on which they act. Equal superpositions of these states have been shown to have Heisenberg limited sensitivity in estimating the phase of a global oscillator \cite{Bollinger1996,Leibfried2004}.

To hold a more general discussion on the use of QECC for the purpose of metrology we turn to the stabilizer formalism \cite{Gottesman}. Stabilizer QECC's are spanned by the $+1$ degenerate eigenstates of an Abelian subgroup of the Pauli group of operators, acting in the full Hilbert space of $n$ spins. For simplicity we limit the discussion here to QECC's that encode a single probe-qubit, although the following ideas can be easily extended to codes that encode more probe-qubits. With a distance between the code words, i.e. the minimum weight of the Pauli operator that converts one code word into another, being $d$, these $[[n,1,d]]$ codes correct for the errors ${E_j}$, all of weight $t<(d-1)/2$. The stabilizer subgroup is generated by the $n-1$ code stabilizers, $\{ {S_i} \}$. The errors, $\{E_j\}$ that the code detects satisfy the condition of anti-commuting with at least one of the code stabilizers. The errors that the code corrects satisfy the condition that for every $i,j$, $E^{\dagger}_{i}E_{j}$ is either in the stabilizer group or anti-commutes with at least one of the stabilizer group generators. Operators that the code cannot detect or correct are in the normalizer to the code, i.e. commute with all the code stabilizer, but are not included in the stabilizer group. These operators result in evolution of states within the code subspace. If these operators act in an uncontrolled fashion they result in decoherence of states in the QECC. In quantum computing, the controlled action of these operators is used to perform logical operations. In the context of quantum metrology these operators result in the measurement signal. Here we assume that these operators (hereafter referred to as the signal operators) are part of the  Hamiltonian and act continuously with a given coupling strength. We investigate the uncertainty with which this coupling strength can be estimated by measuring the protected qubit evolution. Again, the signal operators can act uncontrollably, and their coupling strength estimated via a quantitative study of qubit decoherence. Alternatively, these operators can act in a controlled fashion and measurement will be done via phase estimation in the QECC. Here we focus on the later case.

Suppose a Hamiltonian, $H = \hbar\xi \tilde{X}$, generates an X-rotation in a QECC. The $\pm 1$ code eigenstates of $\tilde{X}$ are $|\pm \tilde{X}\rangle$. We would like to estimate $\xi$ based on the resulting rotation of an encoded state within the QECC. To be most sensitive to an X-rotation we initialize our probe in a state orthogonal to the rotation eigenstates; e.g. $|\phi(t=0)\rangle = \frac{1}{\sqrt{2}}(|\tilde{X}\rangle + |-\tilde{X}\rangle)$. In the absence of external noise, $|\phi\rangle$ evolves as $|\phi(t=\tau)\rangle = \frac{1}{\sqrt{2}}(e^{i\xi\tau}|\tilde{X}\rangle + e^{-i\xi\tau}|-\tilde{X}\rangle)$. We subsequently measure the protected qubit in the basis $|\pm\tilde{Z}\rangle = \frac{1}{\sqrt{2}}(|\tilde{X}\rangle \pm |\tilde{X}\rangle)$. The uncertainty of estimating $\xi$ is calculated using the Cramer-Rao bound, $\Delta\xi = \Delta\tilde{Z}/(\partial\langle\tilde{Z}\rangle/\partial\xi)$. Here $\langle\tilde{Z}\rangle$ is the expectation value of $\tilde{Z} = |+\tilde{Z}\rangle\langle + \tilde{Z}| - |-\tilde{Z}\rangle\langle - \tilde{Z}|$, and $\Delta\tilde{Z} = \sqrt{\langle\tilde{Z}^2\rangle - \langle\tilde{Z}\rangle^2}$, is the uncertainty in measuring $\tilde{Z}$ due to quantum projection noise. An easy calculation shows that $\Delta\xi = 1/2\tau$. As expected, a single measurement of a protected qubit yields an uncertainty in parameter estimation that corresponds to that of a single qubit. The fact that the code involves $n$ physical qubits does not reduce the uncertainty in parameter estimation; i.e. we lose sensitivity as compared with the SQL of $n$ uncorrelated qubits.

The protection offered by QECC is largely provided by the redundancy of physical qubits used to encode every logical qubit. This redundancy results in a redundancy in the number of, physically-different, ways a certain rotation can be implemented on a protected qubit. Back to the three qubit repetition code above, when acting on the code basis states,  $|\uparrow\uparrow\uparrow\rangle$ and $|\downarrow\downarrow\downarrow\rangle$, the operators $ZII$, $IZI$, $IIZ$ and $ZZZ$ all introduce a $\pi$ phase shift between them. All four operators therefore can be equally used to generate a $\tilde{Z}$ rotation in the protected subspace. To see how many physically different ways a given rotation can be implemented in a given code, we notice that if $A$ and $B$ are linearly independent and implement the same rotation on a code word, i.e. $A|\psi\rangle = B|\psi\rangle$, then necessarily $A = BC$ where $C$ is different from the identity and $C|\psi\rangle = |\psi\rangle$. In other words, $A$ can be written as $B$ times an operator from the stabilizer group. The number of different operators which implement the same rotation within the protected subspace can therefore be calculated by construction: starting from a given rotation $A$, all other physical implementations of this rotation can be written as $AS_i$ where $S_i$ is a member of the stabilizer group (notice that all members of the stabilizer group square to the identity). For an $[[n,1,d]]$ code, the stabilizer group has $n-1$ generators, and therefore the order of the stabilizer group (excluding its center) is $n(n-1)/2$. The number of different physical ways to implement a given rotation is therefore $n(n-1)/2+1$. Correspondingly, for a three-qubit code there are four different ways to implement each rotation, for a five-qubit code there are eleven different ways etc.

Now consider a Hamiltonian with a linear combination of $M$ operators, all generating the same X-rotation in the protected subspace,  $H = \hbar\xi \sum_{i=1}^{M}\tilde{X_i}$. The evolution of $|\psi\rangle$ will be $M$ times faster, $|\phi(t=\tau)\rangle = \frac{1}{\sqrt{2}}(e^{iM\xi\tau}|\tilde{X}\rangle + e^{-iM\xi\tau}|-\tilde{X}\rangle)$. Measuring again in the $\tilde{Z}$ basis, $\xi$ can be estimated with an uncertainty of $\Delta\xi = 1/2M\tau$. The $M$-fold faster evolution of the protected state results in a smaller uncertainty by a factor $M^{-1}$. With the large $n(n-1)/2+1$ redundancy in the number of different operators that implement the same rotation, it seems that the gain in the estimation precision of $\xi$ can be very large and even exceed the Heisenberg limit.

One  assumption we made in the analysis above is that all operators appear with exactly the same strength in the Hamiltonian. This is not necessarily a natural situation. It is however natural in cases where the different operators differ by cyclic permutations. In this case, they represent the same physical action; e.g. action of an external field or interactions among spins, only involving different qubits in the code. In terms of the Pauli operators involved, cyclic permutations will appear as a collective shift, right or left, of all operators. For all operators with periodicity of $n$ there would be $n$ different such permutations.

Lets examine two simple examples. The three qubit code mentioned above is a cyclic code with the stabilizer generators, $S_1 = ZZI$ and $S_2 = IZZ$. It corrects against the three single-qubit rotations, $XII$, $IXI$, and $IIX$. Z-rotations in this code are generates by $\tilde{Z}_1 = ZII$, $\tilde{Z}_2 = \tilde{Z}_1\cdot S_1 = IZI$, and $\tilde{Z}_3 = \tilde{Z}_2\cdot S_3 = IIZ$; a permutation of the single qubit Z-rotation between the three different probes. The Hamiltonian $H = \hbar\xi \sum_{i=1}^{3}\tilde{Z_i}$ can be measured with an uncertainty of $\Delta\xi = 1/6\tau$ after one measurement. As noted in \cite{Bollinger1996}, this will enable the Heisenberg limited estimation of a phase difference between a global oscillator and an array of probes with an identical transition frequency. Here, the assumption of identical coupling in the Hamiltonian relies on the probe transitions being degenerate. This assumption can be reasonably satisfied in experiments using identical probes, such as atoms or ions, and is limited by their inhomogenous broadening.

A second example is the five qubit cyclic code which satisfies the Hamming bound for single qubit error-correction by non-degenerate codes. The five-qubit code corrects against any single-qubit rotation and its generators are the five different cyclic permutations of $XZZXI$. Correspondingly, the five cyclic permutations of $YZYII$ all implement Z-rotations within the code subspace. Given an identical three-qubit coupling of the form above acting on all nearest-neighbor probe triplets in the code (assuming circular boundary conditions), $H = \hbar\xi (YZYII + IYZYI + IIYZY + YIIYZ + ZYIIY)$, the coupling strength $\xi$ can be estimated with Heisenberg limited uncertainty, $\Delta\xi = 1/10\tau$. Here again, the different Z-rotations in the code are due to the same physical process only acting on different probe triplets. Since the physics is the same, control over parameters in the experiment, e.g. the distance between probes, can reduce the inhomogeneity in the coupling of different terms.

How truly beneficial is the above usage of QECC's to the metrology end? In arrays of interacting spins, most Hamiltonian terms, and therefore also the signals that are typically measured, are of a form similar to either $\vec{J}\cdot\vec{n}$ or $\vec{J}^2$, where $\vec{J} = \sum_i \vec{\sigma}_i$ and $\vec{\sigma}_i$ is a Cartesian vector of the three Pauli operators of probe $i$. These two operators correspond to either independent action of an external field on different probes or bipartite interactions between different probes. The operators above are of weight one and two and therefore can only be detected, in the sense outlined above, in distance one or two codes respectively. In order to correct any single qubit error, however, a minimum of distance three code is needed. Distance two codes can detect any single qubit error or correct for a subset of single qubit errors. It would still make sense to use such codes for metrology in situations where the noise is limited to one primary noisy Pauli operation; e.g. dephasing or uncorrelated probe-qubit decay, or in situations where noise is limited to a small number of probes in the code, as in the erasure channel. Weight three terms in the Hamiltonian, which can be directly measured in QECC's that protect against any single qubit rotation, physically correspond to direct three probe-qubit interaction. These tripartite interactions are typically weak when occurring naturally, and are certainly hard to engineer. On the other hand, this also means that, when they do occur, they are harder to measure without the use of QECC's.

The effect of detectable, yet uncorrectable, errors is as disruptive to metrology as it is to quantum computing. When trying to detect the evolution under some signal operator, the occurrence of such an error cannot be differentiated from an occurrence of a correctable error together with legitimate evolution of the probe qubit under a signal operator in the protected subspace, leading to measurement error. This can be a real liability, because it implies that the coupling of the signal operators to the probe qubit have to be larger than that of detectable and uncorrectable errors, whereas the former are of higher weight than the latter. On the other hand, in situations where the signal operators are completely negligible, QECC's can be used to estimate the coupling strength of detectable errors by quantifying correlated events of error syndrome detection together with probe rotation.

In conclusion here we analyzed the use of QECC's for the purpose of weak signal measurement. We have shown that the coupling strength of the same operators that are used as single qubit rotations in quantum computing, can be estimated by measuring rotations of the protected qubit. Furthermore, the redundancy in the number of physically different ways the same rotation can be implemented in QECC's leads to a possible reduction in the estimation uncertainty. In codes in which the same rotation is implemented by a cyclic permutation of the same Pauli operators between the probe-qubits in the code, i.e. by the same physical operation only acting on different probes in the code, the estimation of their common coupling strength can be performed with Heisenberg limited sensitivity. The long coherence times that are, in principle, reachable by using QECC's, together with the possibility of measuring certain signals with sub SQL uncertainty, suggests that these protocols can be used as a very powerful metrological tool.

We thank S. Kotler, N. Akerman, D. Gottesman and A. Retzker for fruitful discussions. This research was supported by the Israeli Science Foundation. During the preparation of this manuscript we learned of a similar suggestion for the use of QECC's for metrology \cite{Alex}.

\end{document}